\documentclass{article}%
\usepackage{amsmath}
\usepackage{amsfonts}
\usepackage{amssymb}
\usepackage{graphicx}%
\providecommand{\U}[1]{\protect\rule{.1in}{.1in}}
\begin{document}
\begin{titlepage}
\vspace{.3cm} \vspace{1cm}
\begin{center}
\baselineskip=16pt \centerline{\Large\bf Massive Hermitian Gravity } \vspace{2truecm} \centerline{\large\bf Ali H.
Chamseddine$^{1,2}$\ , \ Viatcheslav Mukhanov$^{3,4}$\ \ } \vspace{.5truecm}
\emph{\centerline{$^{1}$Physics Department, American University of Beirut, Lebanon}}
\emph{\centerline{$^{2}$I.H.E.S. F-91440 Bures-sur-Yvette, France}}
\emph{\centerline{$^{3}$Theoretical Physics, Ludwig Maxmillians University,Theresienstr. 37, 80333 Munich, Germany }}
\emph{\centerline{$^{4}$LPT de l'Ecole Normale Superieure, Chaire Blaise Pascal, 24 rue Lhomond, 75231 Paris cedex, France}}
\end{center}
\vspace{2cm}
\begin{center}
{\bf Abstract}
\end{center}
Einstein-Strauss Hermitian gravity was recently formulated as a gauge theory where the tangent group is taken to be
the pseudo-unitary group instead of the orthogonal group.  A Higgs mechanism for massive gravity was also formulated.
We generalize this construction to obtain massive Hermitian gravity with the use of a complex Higgs multiplet. We show
that both the graviton and antisymmetric tensor acquire the same mass. At the linearized level, the theory is ghost free
around Minkowski background and describes a massive graviton with five degrees of freedom and an antisymmetric
field with three degrees of of freedom. We determine the strong coupling scales for these degrees of freedom and
argue that the potential nonlinear ghosts, if they exist, have to decouple from the gravitational degrees of freedom in strong
coupling regime.
\end{titlepage}

\section{\bigskip Introduction}

Hermitian gravity is based on a Hermitian metric tensor unifying gravity with
an antisymmetric tensor. It was first formulated by Einstein \cite{Einstein}
and then by Einstein and Strauss \cite{ES} in the hope of unifying gravity
with electromagnetism based on a geometrical construction with a Hermitian
affine connection. Schr\"{o}dinger has shown that Hermitian gravity is
equivalent to a theory of gravity with a non-symmetric metric tensor
\cite{Schr}. There exists variations of this theory depending on whether a
first or second order formulation is used. A\ systematic study of all these
models was undertaken by Damour, Deser and McCarthy \cite{DD}, who have shown
that these suffer either from appearance of ghost states or impose
unacceptable constraints on the curvature tensor. They arrived at a no-go
theorem for all these models, which however, could be evaded by adding a mass
term to the antisymmetric tensor, or a cosmological constant formed from the
determinant of the Hermitian metric. Adding a mass term to the antisymmetric
field could not be written using the Hermitian metric only, and thus it is not
geometrical. This contradicts the fundamental assumption of Hermitian gravity
that all geometric invariants must be expressed in terms of the Hermitian metric.

Recently much progress was made in formulating a consistent theory of massive
gravity where the graviton acquires mass through the Higgs mechanism involving
four scalar fields \cite{tHooft, ChM1, ChM2, ChM3, ChM4}. The vacuum
expectation values of these fields cause the excitations of three of the four
scalar fields to be absorbed by the metric thus leading to a massive graviton
with five degrees of freedom. The fourth scalar, a potential ghost degree of
freedom, is non-dynamical in the linear approximation on the Minkowski
background for the Fierz-Pauli mass term \cite{pauli}. The potential nonlinear
ghost \cite{boul}, if exists, is in strong coupling regime above Vainshtein
energy scale \cite{Vainshtein} and, hence, harmless for gravity \cite{ChM3}.
It is then natural to consider whether the Higgs mechanism generalizes to
Hermitian gravity. The aim would then be to give a mass to the antisymmetric
field through spontaneous symmetry breaking mechanism with the four complex
scalar fields $z^{A}$ involved. In analogy with the case of real scalar fields
with global $SO(1,3)$ symmetry, the complex scalar fields must be taken to
have a global $U(1,3)$ symmetry which must be imposed on the couplings to the
Hermitian metric.

In this letter, we shall construct a model of massive Hermitian gravity
coupled to four complex scalar fields $z^{A}.$ First, we will briefly review a
derivation of the Hermitian gravity action based on gauging of the $U(1,3)$
symmetry and promoting it to be the tangent group of the manifold \cite{Ch,
ChMherm}. Then we will introduce four complex scalar fields and show how these
fields, when acquire vacuum expectation values, generate the same masses for
both the graviton and the antisymmetric tensor. Finally we determine
non-covariantly the physical degrees of freedom of the antisymmetric field and
find the strong coupling scales for them.

\section{ Hermitian gravity and the $U(1,3)$ tangent group}

The action for Hermitian gravity is most easily constructed via imposing a
local $U(1,3)$ symmetry which is identified with the tangent group of the
four-dimensional manifold. The main fields are then a complex vierbein
$e_{a}^{\mu}$ and the connections $\omega_{\mu b}^{a},$ $\Gamma_{\rho\mu}%
^{\nu}\left(  g\right)  ,$ constrained by the metricity conditions
\cite{ChMherm}:
\begin{equation}
0=\nabla_{\mu}e_{a}^{\nu}=\partial_{\mu}e_{a}^{\nu}+\omega_{\mu a}^{\quad
b}e_{b}^{\nu}+\Gamma_{\rho\mu}^{\nu}\left(  g\right)  e_{a}^{\rho}.
\label{metricity}%
\end{equation}
These sixty four complex conditions could be solved, perturbatively, to
determine the 64 (anti) Hermitian spin-connections $\left(  \omega_{\mu
b}^{\quad c}\right)  ^{\ast}\eta_{ca}=-\omega_{\mu a}^{\quad c}\eta_{cb}$ and
the 64 Hermitian connections $\left(  \Gamma_{\mu\rho}^{\nu}\right)  ^{\ast
}=\Gamma_{\rho\mu}^{\nu}$ in terms of $e_{a}^{\mu}.$ The curvature tensor is
identified with the field strength
\begin{equation}
R_{\mu\nu a}^{\quad\quad b}=\partial_{\mu}\omega_{\nu a}^{\quad b}%
-\partial_{\nu}\omega_{\mu a}^{\quad b}+\omega_{\mu a}^{\quad c}\omega_{\nu
c}^{\quad b}-\omega_{\nu a}^{\quad c}\omega_{\mu c}^{\quad b}, \label{1}%
\end{equation}
which admits two possible contractions$:$%
\begin{equation}
R=\,e_{b}^{\mu}R_{\mu\nu a}^{\quad\quad b}e^{\nu a}, \label{2a}%
\end{equation}%
\begin{equation}
\tilde{R}=g^{\mu\nu}R_{\mu\nu a}^{\quad\quad a}, \label{2}%
\end{equation}
where
\begin{equation}
g^{\mu\nu}=e_{a}^{\mu}e^{\nu a},\text{ \ \ }e^{\mu a}=\left(  e_{b}^{\mu
}\right)  ^{\ast}\eta^{ab},\text{ \ \ }e_{a}^{\mu}e_{\nu}^{a}=\delta_{\nu
}^{\mu} \label{3}%
\end{equation}
We have shown in \cite{ChMherm}, that using the constraint (\ref{metricity})
the curvatures above can be expressed in terms of $\Gamma_{\rho\mu}^{\nu}:$%
\begin{equation}
R\left(  \omega\right)  =\eta^{ac}e_{c}^{\mu\ast}R_{\mu\nu a}^{\quad
\,\,\,\,b}\left(  \omega\right)  e_{b}^{\nu}=-\eta^{ac}e_{c}^{\mu\ast
}R_{\,\,\,\rho\mu\nu}^{\nu}\left(  \Gamma\right)  e_{a}^{\rho}=g^{\rho\mu
}R_{\,\,\,\rho\nu\mu}^{\nu}\left(  \Gamma\right)  =R\left(  \Gamma\right)
,\nonumber
\end{equation}
where
\begin{equation}
R_{\hspace{0.03in}\hspace{0.03in}\rho\mu\nu}^{\sigma}\left(  \Gamma\right)
=\partial_{\mu}\Gamma_{\rho\nu}^{\sigma}-\partial_{\nu}\Gamma_{\rho\mu
}^{\sigma}+\Gamma_{\kappa\mu}^{\sigma}\Gamma_{\rho\nu}^{\kappa}-\Gamma
_{\kappa\nu}^{\sigma}\Gamma_{\rho\mu}^{\kappa}. \label{4}%
\end{equation}
The generalization of the Einstein action is given then by
\begin{equation}
S_{E}=-\frac{1}{2}%
%TCIMACRO{\dint }%
%BeginExpansion
{\displaystyle\int}
%EndExpansion
d^{4}x\left\vert \det e_{\mu}^{a}\right\vert R, \label{5}%
\end{equation}
and depends on metric $g_{\mu\nu}$ only (we use the units where $8\pi G=1$)

\section{Higgs for gravity}

To give mass to the graviton and antisymmetric complex part of the metric we
will introduce the four complex scalar fields $z^{A}$ and construct the
induced \textquotedblleft Hermitian metric\textquotedblright\
\begin{equation}
H_{\text{ }B}^{A}=g^{\mu\nu}\partial_{\mu}z^{A}\partial_{\nu}z_{B}=\left(
H_{\text{ }C}^{D}\right)  ^{\ast}\eta_{BD}\eta^{AC},\label{6}%
\end{equation}
where we have defined
\begin{equation}
z_{A}=\eta_{AB}\left(  z^{B}\right)  ^{\ast}.\label{7}%
\end{equation}
It is straightforward then to write the analogue of the Fierz-Pauli mass term
in powers of $\bar{H}_{\text{ }B}^{A}$, defined by
\begin{equation}
\bar{H}_{\text{ }B}^{A}=H_{\text{ }B}^{A}-\delta_{B}^{A}.\label{8}%
\end{equation}
The action for the complex scalar fields providing us with the mass term for
gravity becomes
\begin{equation}
S_{z}=\frac{m^{2}}{8}\int d^{4}x\sqrt{-g}\left(  \bar{H}^{2}-\bar{H}_{\text{
}B}^{A}\bar{H}_{\text{ }A}^{B}+O\left(  \bar{H}^{3}\right)  \right)
.\label{fierz}%
\end{equation}
where by $O\left(  \bar{H}^{3}\right)  $ we have denoted all possible higher
order extensions of the Fierz-Pauli term, which do not influence the linear
propagator for massive graviton on the symmetry broken background. An elegant
non-linear extension of the action for complex scalar fields  $\left(
\ref{fierz}\right)  $ will be given in the appendix. The vacuum solution of
the full action given by the sum of (\ref{5}) and (\ref{fierz}) is%
\begin{equation}
g^{\mu\nu}=\eta^{\mu\nu},\text{ \ }z^{A}=x^{A}.\text{\ \ }\label{10}%
\end{equation}
Expanding $z^{A}$ around this vacuum solution, we write
\begin{equation}
z^{A}=x^{A}+\chi^{A}+i\psi^{A},\label{11}%
\end{equation}
while for the metric $g^{\mu\nu}$ we have
\begin{equation}
g^{\mu\nu}=\eta^{\mu\nu}+h^{\mu\nu}+iB^{\mu\nu},\label{12}%
\end{equation}
with $B^{\mu\nu}$ being antisymmetric. Similarly $\bar{H}_{\text{ }B}^{A},$
which is Hermitian, can be decomposed in terms of a real symmetric part and an
imaginary antisymmetric part%
\begin{equation}
\bar{H}_{\text{ }B}^{A}=\bar{h}_{\text{ }B}^{A}+i\bar{B}_{\text{ }B}%
^{A},\label{13}%
\end{equation}
where $\bar{h}^{AB}=\bar{h}^{BA}$ and $\bar{B}^{AB}=-\bar{B}^{BA}$ and the
indices are raised and lowered with the Minkowski metric $\eta_{AB}$.
Substituting the expansions (\ref{11}), (\ref{12}) into the definition of
$\bar{H}_{\text{ }B}^{A}$ we find%
\begin{equation}
\bar{h}_{AB}=h_{AB}+\partial_{A}\chi_{B}+\partial_{B}\chi_{A}+O(\left(
\partial\chi\right)  ^{2},...),\label{14}%
\end{equation}%
\begin{equation}
\bar{B}_{AB}=B_{AB}-\partial_{A}\psi_{B}+\partial_{B}\psi_{A}+O(\partial
\chi\partial\psi,...),\label{15}%
\end{equation}
where we have denoted by $O(\left(  \partial\chi\right)  ^{2},...)$ the higher
order terms in perturbations, the explicit form of which will not be needed
here. Notice that $\bar{h}_{AB}$ is invariant with respect infinitesimal
diffeomorphism transformations and the antisymmetric field $\bar{B}_{AB}$ is
invariant with respect to the infinitesimal gauge transformations%
\begin{equation}
B_{AB}\rightarrow B_{AB}+\partial_{A}\zeta_{B}-\partial_{B}\zeta_{A},\text{
\ }\psi_{A}\rightarrow\psi_{A}+\zeta_{A}\text{\ .}\label{16}%
\end{equation}
Substituting (\ref{13}) into (\ref{fierz}) we can rewrite the action for the
scalar fields up to quadratic order as%
\begin{equation}
S_{z}=\frac{m^{2}}{8}\int d^{4}x\sqrt{-g}\left(  \bar{h}^{2}-\bar{h}^{AB}%
\bar{h}_{AB}-\bar{B}^{AB}\bar{B}_{AB}\right)  .\label{17}%
\end{equation}

Next we expand the Einstein action (\ref{5}) up to second order in
perturbations. Using the equivalence of the expressions for curvature in terms
of spin-connections $\omega_{\mu a}^{\quad b}$ to that in terms of the
Hermitian connections $\Gamma_{\mu\nu}^{\rho}\left(  g\right)  $, we can solve
the equation
\begin{equation}
\partial_{\mu}g^{\nu\rho}+\Gamma_{\sigma\mu}^{\nu}g^{\sigma\rho}%
+\Gamma_{\sigma\mu}^{\rho}g^{\nu\sigma}=0, \label{18}%
\end{equation}
to determine $\Gamma_{\mu\nu}^{\rho}\left(  g\right)  $ perturbatively in
terms of powers of $h^{\mu\nu}$ and $B^{\mu\nu}.$ To the first order we have
\begin{equation}
\Gamma^{\rho\mu\nu\left(  1\right)  }=-\frac{1}{2}\left(  \partial^{\mu
}\left(  h^{\nu\rho}+iB^{\nu\rho}\right)  -\partial^{\nu}\left(  h^{\rho\mu
}+iB^{\rho\mu}\right)  +\partial^{\rho}\left(  h^{\mu\nu}+iB^{\mu\nu}\right)
\right)  , \label{19}%
\end{equation}
where the indices are raised and lowered with the Minkowski metric. This can
be used back in the constraint equation to find $\Gamma^{\rho\mu\nu\left(
2\right)  }$ to second order. The gravitational action to second order is then
given by
\begin{align}
S_{E}=  &  \frac{1}{8}%
%TCIMACRO{\dint }%
%BeginExpansion
{\displaystyle\int}
%EndExpansion
d^{4}x\left[  \left(  \partial^{A}h^{BC}\partial_{A}h_{BC}+2\partial
_{B}h\partial_{A}h^{AB}-2\partial_{A}h^{AB}\partial^{C}h_{CB}-\partial
_{A}h\partial^{A}h\right)  \right. \nonumber\\
&  \qquad\qquad\left.  +\left(  \partial^{A}B^{BC}\partial_{A}B_{BC}%
-2\partial^{C}B^{AB}\partial_{A}B_{CB}\right)  \right]  \label{20}%
\end{align}
This action is invariant with respect to both the diffeomorphism and gauge
transformations, respectively,%
\begin{equation}
h_{AB}\rightarrow h_{AB}+\partial_{A}\xi_{B}+\partial_{B}\xi_{A},\text{
\ }B_{AB}\rightarrow B_{AB}+\partial_{A}\zeta_{B}-\partial_{B}\zeta
_{A}\text{.\ } \label{21}%
\end{equation}
Hence we can replace $h_{AB}$ and $B_{AB}$ in the gravitational part of the
action by their gauge invariant combinations with the scalar fields, $\bar
{h}_{AB}$ and $\bar{B}_{AB},$ correspondingly. The field $\bar{h}_{AB}$ then
satisfy the same linear equations as massive graviton with five degrees of
freedom. The massive gravity and its nonlinear extensions were studied in
details in \cite{ChM1, ChM2, ChM3, ChM4} and therefore we concentrate here
only on the antisymmetric massive field, the action for which becomes%
\begin{equation}
S_{\bar{B}}=\frac{1}{8}%
%TCIMACRO{\dint }%
%BeginExpansion
{\displaystyle\int}
%EndExpansion
d^{4}x\left(  \partial^{A}\bar{B}^{BC}\partial_{A}\bar{B}_{BC}-2\partial
^{C}\bar{B}^{AB}\partial_{A}\bar{B}_{CB}-m^{2}\bar{B}^{AB}\bar{B}_{AB}\right)
. \label{22}%
\end{equation}
The equations of motion for $\bar{B}_{AB}$ are
\begin{equation}
\left(  \partial^{2}+m^{2}\right)  \bar{B}_{AB}-\partial_{A}\partial^{C}%
\bar{B}_{CB}-\partial_{B}\partial^{C}\bar{B}_{AC}=0. \label{23}%
\end{equation}
They describe massive field with three degrees of freedom. Remarkably,
$\bar{B}_{AB}$ is exactly the same combination of fields worked out by Kalb
and Ramond in \cite{KR}, where they used St\"{u}ckelberg method to introduce
fake gauge invariance for the auxiliary fields (corresponding here to
$\psi^{A}$ ) and showed that $\bar{B}_{AB}$ has two degrees of freedom from
the transverse components of $\psi^{A}$ plus one degree from the longitudinal
part of $B_{AB}$.

To demonstrate this explicitly and to determine the strong coupling scales for
the different degrees of freedom we will express the action (\ref{22})
entirely in terms of physical degrees of freedom and find when they enter the
strong coupling regime.

\section{Physical degrees of freedom and strong coupling scales}

Let us first rewrite the action (\ref{22}) explicitly separating space and
time components in $\bar{B}_{AB}:$%
\begin{align}
S_{\bar{B}}  &  =\frac{1}{8}%
%TCIMACRO{\dint }%
%BeginExpansion
{\displaystyle\int}
%EndExpansion
d^{4}x\left[  \dot{B}_{ik}^{2}+2\dot{B}_{ik}\left(  B_{i,k}-B_{k,i}\right)
+\left(  B_{i,k}-B_{k,i}\right)  ^{2}\right. \nonumber\\
&  \left.  -B_{ik,j}^{2}-2B_{ik,j}B_{ji,k}+2m^{2}B_{i}^{2}-m^{2}B_{ik}%
^{2}\right]  , \label{24}%
\end{align}
where dot denotes the derivative with respect to time, indices $i,k,...$ take
the values $1,2,3$ and comma denotes the derivative with respect to the
corresponding spatial coordinate. We have also introduced the following
notations:
\begin{equation}
\bar{B}_{0i}\equiv B_{i},\qquad\bar{B}_{ik}\equiv B_{ik},
\end{equation}
and assumed summation over repeated indices. Next we define the vector $A_{l}$
dual to the antisymmetric tensor $B_{ik}$, so that,%
\begin{equation}
B_{ik}=\varepsilon_{ikl}A_{l}, \label{25}%
\end{equation}
and decompose the 3-vectors $A_{l}$ and $B_{i}$ into transverse and
longitudinal parts%
\begin{equation}
A_{l}=\frac{\varphi_{,l}}{\sqrt{-\Delta}}+A_{l}^{\left(  T\right)  },\text{
\ \ \ \ }B_{i}=\mu_{,i}+B_{i}^{\left(  T\right)  }, \label{26}%
\end{equation}
where $\Delta$ is the Laplacian and the transverse components satisfy the
conditions $A_{l,l}^{\left(  T\right)  }=0,B_{i,i}^{\left(  T\right)  }=0.$
Substituting (\ref{25}) and (\ref{26}) into (\ref{24}) the action reduces to%
\begin{align}
S_{\bar{B}}  &  =\frac{1}{4}%
%TCIMACRO{\dint }%
%BeginExpansion
{\displaystyle\int}
%EndExpansion
d^{4}x\left[  \left(  \dot{\varphi}^{2}-\varphi_{,i}\varphi_{,i}-m^{2}%
\varphi^{2}\right)  +m^{2}\mu_{,i}\mu_{,i}+\dot{A}_{i}^{\left(  T\right)
}\dot{A}_{i}^{\left(  T\right)  }\right. \nonumber\\
&  +2\varepsilon_{ikl}B_{i,k}^{\left(  T\right)  }\dot{A}_{l}^{\left(
T\right)  }+B_{i,k}^{\left(  T\right)  }B_{i,k}^{\left(  T\right)  }%
-m^{2}A_{i}^{\left(  T\right)  }A_{i}^{\left(  T\right)  }+m^{2}B_{i}^{\left(
T\right)  }B_{i}^{\left(  T\right)  }. \label{28}%
\end{align}
Variation of this action with respect to $\mu$ and $B_{i}^{\left(  T\right)
}$ give us the constraints
\begin{equation}
\Delta\mu=0,\text{ \ \ }\varepsilon_{ikl}\dot{A}_{l,k}^{\left(  T\right)
}\text{\ }+\Delta B_{i}^{\left(  T\right)  }-m^{2}B_{i}^{\left(  T\right)
}=0, \label{29}%
\end{equation}
from which it follows, that%
\begin{equation}
\mu=0,\text{ \ \ }B_{i}^{\left(  T\right)  }=\frac{\varepsilon_{ikl}\dot
{A}_{l,k}^{\left(  T\right)  }}{-\Delta+m^{2}}. \label{30}%
\end{equation}
Substituting these expressions into (\ref{28}), the action becomes
\begin{equation}
S_{\bar{B}}=\frac{1}{4}%
%TCIMACRO{\dint }%
%BeginExpansion
{\displaystyle\int}
%EndExpansion
d^{4}x\left[  \left(  \dot{\varphi}^{2}-\varphi_{,i}\varphi_{,i}-m^{2}%
\varphi^{2}\right)  +\left(  \dot{A}_{i}^{\left(  T\right)  }\frac{m^{2}%
}{-\Delta+m^{2}}\dot{A}_{i}^{\left(  T\right)  }-m^{2}A_{i}^{\left(  T\right)
}A_{i}^{\left(  T\right)  }\right)  \right]  . \label{31}%
\end{equation}
The three physical degrees of freedom (one pseudo-scalar $\varphi$ and two
independent transverse components of pseudo-vector $A_{i}^{\left(  T\right)
}$) satisfy the following equations
\begin{equation}
\left(  \partial^{2}+m^{2}\right)  \varphi=0,\text{ \ \ }\left(  \partial
^{2}+m^{2}\right)  A_{i}^{\left(  T\right)  }=0. \label{32}%
\end{equation}
The last term in the action is proportional to the mass and therefore when the
mass $m$ vanishes $A_{i}^{\left(  T\right)  }$ drops out from the action and
in this limit the antisymmetric field $B_{\alpha\beta}$ describes a massless
pseudo-scalar with only one degree of freedom. This is not surprising because
as one can easily see from (\ref{24}), three $B_{0i}$ components of the
antisymmetric metric are not dynamical and the gauge symmetry (\ref{21}),
involving only two transverse components of\ $\zeta_{i}$ removes two degrees
of freedom in $B_{ik}.$ When we couple gravity to the scalar fields $\psi
_{B},$ which in the absence of $B_{\mu\nu}$ are described by the
\textquotedblleft Maxwell action\textquotedblright\
\begin{equation}
-\frac{m^{2}}{8}\int d^{4}x\bar{B}^{AB}\bar{B}_{AB}, \label{33}%
\end{equation}
with $\bar{B}_{AB}=\partial_{B}\psi_{A}-\partial_{A}\psi_{B},$ the two
physical degrees of freedom of $\psi_{B}$ are absorbed by the antisymmetric
metric, which thus acquires three degrees of freedom.

The scalar and vector degrees of freedom become strongly coupled at different
scales. To determine these scales let us consider the plane wave with the
wavelength $\lambda.$ We first note that the scalar $\varphi$ enters
(\ref{31}) with canonical normalization. Hence, the minimal quantum
fluctuations of this field at the length-scale $\lambda$ are of order
$\delta\varphi_{\lambda}\simeq1/\lambda$ for $\lambda\ll m^{-1}.$ Because
$\bar{B}\sim\varphi$ (see (\ref{25}), (\ref{26})), we find that the quantum
fluctuations of the antisymmetric field due to the scalar mode become of the
order unity at the Planck scale, where this degree of freedom enters the
strong coupling regime. For the two transverse degrees of freedom
$A_{i}^{\left(  T\right)  }$ the strong coupling scale is larger than the
Planck length. Actually, as follows from (\ref{31}) the canonically normalized
degrees of freedom for these modes are
\[
\sqrt{\frac{m^{2}}{-\Delta+m^{2}}}A_{i}^{\left(  T\right)  }\sim m\lambda
A_{i}^{\left(  T\right)  }%
\]
and therefore the minimal vacuum fluctuations of $m\lambda A_{i}^{\left(
T\right)  }$ decay as $1/\lambda$ for $\lambda\ll m^{-1}.$Hence the amplitude
of the fluctuations of the fields $A_{i}^{\left(  T\right)  }$ itself is of
order
\[
\delta A_{\lambda}^{\left(  T\right)  }\simeq\frac{1}{m\lambda^{2}}%
\]
It then follows that the quantum fluctuations of $\bar{B}\sim\delta
A_{\lambda}^{\left(  T\right)  }$ due to transverse degrees of freedom become
of order unity at scales%
\[
\lambda_{strong}\simeq m^{-1/2}=\frac{1}{m}\left(  \frac{m}{m_{Pl}}\right)
^{1/2}%
\]
For masses much smaller than the Planck mass $m_{Pl}$ this strong coupling
scale is significantly smaller than the inverse mass of the field but larger
than the Planck wavelength by the factor $\left(  m_{Pl}/m\right)  ^{1/2}.$
Thus the transverse modes enter strong coupling regime before the scalar mode.
As a result, the scalar fields which provide mass to these transverse modes
decouple and the antisymmetric field will remain with one scalar degree of
freedom at energy scales above $m^{1/2}.$

In \cite{DD} it was shown that unbroken Hermitian gravity is inconsistent, due
to the coupling of $B_{\mu\nu}$ to the curvature tensor. However, they also
pointed out that the inconsistencies can be avoided by adding a mass term for
the $B_{\mu\nu}$ field. On the other hand, adding non diffeomorphism invariant
mass terms for $B_{\mu\nu}$, destroys the Hermitian symmetry and violates the
diffeomorphism invariance because the antisymmetric field is also a part of
the Hermitian metric. Therefore, \textquotedblleft hard\textquotedblright%
\ introduction of the mass term is not acceptable as it makes the use of a
Hermitian metric pointless and spoils the geometrical nature of Hermitian gravity.

In this paper we have shown that in geometrical Hermitian gravity a mass term
for $B_{\mu\nu}$ can be generated only if the graviton simultaneously acquires
the same mass. By adding four complex scalar fields (corresponding to 8 real
fields), we demonstrated how the gravitational Higgs mechanism can be realized
for Hermitian gravity. Three out of the eight fields are absorbed by the real
symmetric part of the metric thus giving us massive graviton with five degrees
of freedom. Two other fields are absorbed by the antisymmetric part of the
metric making this field massive (with 3 degrees of freedom). The remaining
three scalar fields are non-dymamical at linear level on Minkowski background.
Two of them could be potential non-linear ghosts. However, these potential
ghosts could certainly be dangerous for gravity only in those regions where
the corresponding degrees of freedom are in the weak coupling regime and hence
the perturbative analysis is trustable for them. We have shown that for a
small graviton mass the strong coupling scales are much below the Planck scale
and hence \textquotedblleft the trustable nonlinear ghosts\textquotedblright%
\ are completely harmless even if they would exist.

\section{Appendix: Non-linear extension for Fierz-Pauli mass terms}

It is straightforward to write the analogue of the Pauli-Fierz action
containing various powers of $H_{B}^{A}$. This can be simplified in terms of
the field $\overline{H}_{B}^{A}$ defined by
\begin{equation}
H_{B}^{A}=\delta_{B}^{A}+\overline{H}_{B}^{A}%
\end{equation}
We have shown in reference \cite{ChM4} that in the real case there is a
special simple consistent action that starts with quadratic kinetic terms for
the fields $z^{A}$ instead of the quartic terms normally used . To generalize
this construction, we first define the auxiliary fields $E_{A}^{\mu}$
constrained so that
\begin{equation}
g^{\mu\nu}=E_{A}^{\mu}E^{\nu A}%
\end{equation}
We then define the field
\begin{equation}
S_{B}^{A}=E_{B}^{\mu}\partial_{\mu}z^{A}-\delta_{B}^{A}%
\end{equation}
which depends on the first derivative of $z^{A}$ and is constrained to be
Hermitian%
\begin{equation}
S_{B}^{A}=\left(  S_{A}^{B}\right)  ^{\ast}%
\end{equation}
These constraints could be imposed through the use of Lagrange multipliers.
Thus the 16 complex fields $E_{A}^{\mu}$ are subjected to 32 real constraints,
and could be determined completely, in a perturbative way in terms of
$g^{\mu\nu}$ and $z^{A}.$ The quadratic part of the proposed action in terms
of the field $S_{B}^{A}$ is given by
\begin{equation}%
%TCIMACRO{\dint }%
%BeginExpansion
{\displaystyle\int}
%EndExpansion
d^{4}x\sqrt{g}\left[  \frac{m^{2}}{2!}\delta_{EF}^{AB}S_{A}^{E}S_{B}%
^{F}\right]
\end{equation}
This expression can be rewritten in terms of the induced metric $H_{B}^{A}$ by
using the identity
\begin{align}
S_{C}^{\prime A}S_{B}^{\prime^{C}} &  =E_{C}^{\mu}\partial_{\mu}z^{A}\left(
E_{B}^{\nu}\partial_{\nu}z^{C}\right)  \nonumber\\
&  =E_{C}^{\mu}\partial_{\mu}z^{A}\left(  E_{C}^{\nu}\partial_{\nu}%
z^{B}\right)  ^{\ast}\nonumber\\
&  =E_{C}^{\mu}E^{\nu C}\partial_{\mu}z^{A}\partial_{\nu}z_{B}\nonumber\\
&  =g^{\mu\nu}\partial_{\mu}z^{A}\partial_{\nu}z_{B}\nonumber\\
&  =H_{B}^{A}%
\end{align}
where
\begin{equation}
S_{B}^{\prime A}=S_{B}^{A}+\delta_{B}^{A}%
\end{equation}
Thus we have
\begin{align}
S_{B}^{A} &  =\sqrt{H_{B}^{A}}-\delta_{B}^{A}\nonumber\\
&  =\sqrt{\delta_{B}^{A}+\overline{H}_{B}^{A}}-\delta_{B}^{A}\nonumber\\
&  =\frac{1}{2}\overline{H}_{B}^{A}-\frac{1}{8}\overline{H}_{C}^{A}%
\overline{H}_{B}^{C}+\frac{1}{16}\overline{H}_{C}^{A}\overline{H}_{D}%
^{C}\overline{H}_{B}^{D}+\cdots
\end{align}
There is also unique generalization of the above action that adds terms which
are cubic and quartic in terms of $S_{B}^{A}$%
\begin{equation}%
%TCIMACRO{\dint }%
%BeginExpansion
{\displaystyle\int}
%EndExpansion
d^{4}x\sqrt{g}\left[  \frac{c_{1}}{3!}\delta_{EFG}^{ABC}S_{A}^{E}S_{B}%
^{F}S_{C}^{G}+\frac{c_{2}}{4!}\delta_{EFGH}^{ABCD}S_{A}^{E}S_{B}^{F}S_{C}%
^{G}S_{D}^{H}\right]
\end{equation}
Since $S_{B}^{A}$ is an infinite expansion in terms of $\overline{H}_{B}^{A}$
the action could be expressed in terms of $\overline{H}_{B}^{A}$. The action,
up to quartic terms is given by
\begin{align}
&  -%
%TCIMACRO{\dint }%
%BeginExpansion
{\displaystyle\int}
%EndExpansion
d^{4}x\sqrt{g}\left(  \frac{m^{2}}{2!}\delta_{EF}^{AB}\left(  \frac{1}%
{4}\overline{H}_{A}^{E}\overline{H}_{B}^{F}-\frac{1}{8}\overline{H}_{A}%
^{E}\overline{H}_{C}^{F}\overline{H}_{B}^{C}+\frac{1}{16}\overline{H}_{A}%
^{E}\overline{H}_{C}^{F}\overline{H}_{D}^{C}\overline{H}_{B}^{D}+\frac{1}%
{64}\overline{H}_{C}^{E}\overline{H}_{A}^{C}\overline{H}_{D}^{F}\overline
{H}_{B}^{D}\right)  \right.  \nonumber\\
&  \left.  +\frac{c_{1}}{3!}\delta_{EFG}^{ABC}\left(  \frac{1}{8}\overline
{H}_{A}^{E}\overline{H}_{B}^{F}\overline{H}_{C}^{G}-\frac{3}{32}\overline
{H}_{A}^{E}\overline{H}_{B}^{F}\overline{H}_{D}^{G}\overline{H}_{C}%
^{D}\right)  +\frac{c_{2}}{4!}\frac{1}{16}\delta_{EFGH}^{ABCD}\overline{H}%
_{A}^{E}\overline{H}_{B}^{F}\overline{H}_{C}^{G}\overline{H}_{D}^{H}\right)
\end{align}
This is the same expression obtained in the real case that produce decoupling
of ghosts up to quartic order in perturbation series (compare with equation
(20) in reference \cite{ChM3}).

It would be very interesting to generalize the analysis carried out in this
paper to non-linear terms and on  non-trivial backgrounds. 

\bigskip
%\begin{acknowledgement}

\centerline{\bf Acknowledgments}

The work of AHC is supported in part by the National Science Foundation
0854779. The work of VM is supported by \textquotedblleft Chaire
Internationale de Recherche Blaise Pascal financ\'{e}e par l'Etat et la
R\'{e}gion d'Ile-de-France, g\'{e}r\'{e}e par la Fondation de l'Ecole Normale
Sup\'{e}rieure\textquotedblright, by TRR 33 \textquotedblleft The Dark
Universe\textquotedblright\ and the Cluster of Excellence EXC 153
\textquotedblleft Origin and Structure of the Universe\textquotedblright.

%\end{acknowledgement}

\end{document}